
\font\subtit=cmr12
\font\name=cmr8
\def\ng{\backslash\!\!\!\!D}
\def\rdc#1{{\scriptstyle\sqrt{g(#1)}}}
\input harvmac
\def\MPILMU#1#2#3#4
{\TITLE{MPI-Ph/\number\yearltd-#1}
{LMU-TPW \number\yearltd-#2}{#3}{#4}}
\def\TITLE#1#2#3#4{\nopagenumbers\abstractfont\hsize=\hstitle\rightline{#1}
\vskip 1pt\rightline{#2}
\vskip 1in
\centerline{\subtit #3}
\vskip 1pt
\centerline{\subtit #4}\abstractfont\vskip .5in\pageno=0}%
\MPILMU{71}{26}
{ON THE SCHWINGER MODEL ON RIEMANN SURFACES}{}

\centerline{F{\name RANCO} F{\name ERRARI}\foot{\name This work
is carried out in the framework of the EC Research Programme
``Gauge Theories, Applied Supersymmetry and Quantum Gravity".}}\smallskip
\centerline{\it Max-Planck-Institut f\"ur Physik, P.O. Box 40 12 12,
M\"unchen}\smallskip
\centerline{and}\smallskip
\centerline{\it Sektion Physik der Universit\"at M\"unchen,Theresienstr.
37, 80333 M\"unchen (Fed. Rep. Germany)}
\vskip 1cm
\centerline{ABSTRACT}
\vskip 1cm
{\narrower
In this paper the Schwinger model or two dimensional quantum
electrodynamics is exactly solved on a Riemann surface providing the
explicit expression of the partition function and of the generating
functional of the amplitudes between the fermionic currents.
This offers one of the few examples, if not the only one, in which
it is possible to integrate in an explicit way a gauge field
theory interacting with matter on a Riemann surface.
}
\Date{October 1993}
\newsec { INTRODUCTION}
\vskip 1cm
In this paper we solve the Schwinger model
\ref\sch{J. Schwinger,
{\it Phys. Rev.} {\bf 128} (1962), 2425.} on a closed and orientable
Riemann surface in the absence of topologically
nontrivial gauge fields.
To this purpose, we compute the partition function of the model
and the generating functional of the amplitudes containing fermionic
currents. The method used was developed in the recent papers
\ref\ferqed{F. Ferrari, {\it Class. Q. Grav.},
{\bf 10} (1993), 1065.} and
\ref\ferabe{F. Ferrari, On the Quantization of Abelian Gauge Field
Theories on Riemann Surfaces, Preprint MPI-Ph/93-50, LMU-TPW 93-14.} and
it is able to quantize any
abelian gauge field theory on a Riemann surface.
Here, we restrict ourselves to the Schwinger model, which describes
the quantum electrodynamics in two dimensions or QED$_2$ because, due to
its particular relevance in theoretical physics
\ref\flat{S. Coleman, {\it Phys. Rev.} {\bf D11} (1975), 3026;
S. Donaldson, {\it J. Diff. Geom.}
{\bf 18} (1983), 269;
R. Jackiw, R. Rajaraman, {\it Phys. Rev. Lett.} {\bf 54} (1985), 1219;
L. Faddev, S. Shatashvili, {\it Phys. Lett.} {\bf 183B} (1987), 311;
T. P. Killingback, {\it Phys. Lett.} {\bf 223B} (1989) 357;
A. I. Bocharek, M. E. Schaposhnik, {\it Mod. Phys. Lett.} {\bf A2} (1987),
991; J. Kripfganz, A. Ringwald, {\it Mod. Phys. Lett.} {\bf A5}
(1990), 675.},
\ref\cmpl{J. H. Lowenstein, J. A. Swieca, {\it Ann. Phys.} {\bf 68}
(1961), 172;
A. Z. Capri, R. Ferrari, {\it Nuovo Cim.} {\bf 62A} (1981), 273;
{\it Journ. Math. Phys.} {\bf 25} (1983), 141;
G. Morchio, D. Pierotti, F. Strocchi, {\it Ann. Phys.} {\bf 188}
(1988), 217;
A. K. Raina, G. Wanders,
{\it Ann. of Phys.} {\bf 132} (1981), 404.},
\ref\manton{N. Manton, {\it Ann. Phys.} {\bf 159} (1985), 220; J. E.
Hetrick, Y. Hosotani, {\it Phys. Rev.} {\bf 38} (1988), 2621.},
\ref\jaw{C. Jayewardena, {\it Helv. Phys. Acta} {\bf 61} (1988), 636.},
\ref\Joos{H. Joos, {\it Nucl. phys.} {\bf B17}
(Proc. Suppl.) (1990), 704; {\it Helv. Phys. Acta} {\bf 63} (1990),
670},
\ref\wipf{I. Sachs, A. Wipf, {\it Helv. Phys. Acta} {\bf 65} (1992),
653.}, it provides a significant example of quantum field theory on a
manifold.
The fact that quantization on a manifold is not a trivial procedure
and gives rise to observable phenomena, was discussed in the case of
the Schwinger model in ref. \ferabe.
For instance, the high energy behavior of QED$_2$ strongly depends on
the topology of the space-time.
On a Riemann surface, in fact, the strength of the electromagnetic
forces vanishes at short distances, apart from zero modes which are not
so easy to determine, while on a disk this asymptotic freedom is not
present at all \ferabe.\smallskip
Moreover, it was shown in ref. \ferabe\ that in the path integral of the
Schwinger model defined on a compact manifold without boundary, it is
possible to perform the integration over the gauge fields, obtaining as
a result an effective theory of fermions.
This theory consists in a nonlocal generalization, in the sense of
\ref\moffat{J. W.
Moffat, {\it Phys. Rev.} {\bf D41} (1990), 1177.}, of the massless
Thirring model
\ref\thirring{W. Thirring, {\it Ann. Phys. (N. Y.)} {\bf
3} (1958), 91.}, which will be called here generalized Thirring model or
GTM.
The nonlocality of the GTM together with the property of asymptotic
freedom, make it very interesting by itself.
However, the approach of refs. \ferqed\ and \ferabe,
intended for the quantization of any abelian gauge field theory, was
essentially perturbative and therefore not suitable in order to treat
the GTM, which is integrable.
For this reason, we concentrate here on the Schwinger model, performing
in the path integral also
the integration over the fermionic fields which was
missing in refs. \ferqed\ and \ferabe.
The result is, as we anticipated at the beginning, a nonperturbative
expression of the partition function and of the generating functional of
QED$_2$.
In this way, we obtain one of the few examples, if not the only one, in
which the explicit integration of a gauge field theory
interacting with matter fields is possible on a Riemann surface.
Until now, in fact, the only theories which have been quantized on
Riemann surfaces are either pure gauge field theories or topological field
theories, while the interaction with the matter fields was restricted to
external gauge field without kinetic term
\ref\tomboulis{E. T. Tomboulis, {\it Phys. Lett.} {\bf 198B} (1987),
165; M. Porrati, E. T. Tomboulis, {\it Nucl. Phys.} {\bf B315} (1989),
615.},
\ref\fp{
D. Z. Freedman, K. Pilch, {\it Phys. Lett.}{\bf 213B} (1988), 331.},
\ref\witt{E. Witten, {\it Comm. Math.
Phys.} {\bf 141} (1991), 153; {\it Two
Dimensional Gauge Theories Revisited}, Preprint IASSNS-Hep
92/15.},
\ref\rusakov{
D. J. Gross, W. Taylor IV, Twist and Wilson Loops
in the String Theory of Two Dimensional QCD, Preprint CERN-TH 6827/93,
PUPT-1382, LBL-33767, UCB-PTH-93/09, March 1993;
B. Ye. Rusakov, {\it Mod. Phys. Lett.} {\bf A5}
(1990), 693; A. Migdal, {\it Zh. Esp. Teor. Fiz.} {\bf 69} (1975),
810.}, \ref\froe{J. Fr\"ohlich, On the Construction of Quantized
gauge Fields, in ``Field Theoretical Methods in Particle Physics" (W.
R\"uhl Ed.) Plenum, New York, 1980;
I. L. Buchbinder, S. D. Odintsov, I. L. Shapiro,
{\it Effective Action in Quantum Gravity}, IOP Publishing, Bristol and
Philadelphia, 1992.},
\ref\bt{M. Blau, G. Thompson, {\it Int. Jour. Mod. Phys.}
{\bf A7} (1992), 3781.}, \ref\thom{G. Thompson, 1992 Trieste Lectures on
Topological Gauge Theory and Yang-Mills Theory.},
\ref\fine{D. S. Fine, {\it Comm. Math. Phys.} {\bf 134} (1990), 273;
S. G. Rajeev, {\it Phys. Lett.} {\bf 212B} (1988), 203.}.
The integrability of the Schwinger model, together with its equivalence
with the GTM, make it very interesting not only for string theory, but
also in the study of the effects of a curved space-time on the dynamics
of the fields.
Finally, the quantization of field theory on manifolds can shed some more
light in the quantization of general
relativity, see for example ref. \ref\rov{C. Rovelli, {\it Nucl. Phys.}
{\bf B405} (1993), 797.} on this point.
\smallskip
Apart from the calculation of the propagator of the gauge fields, which
has been carefully discussed in ref. \ferqed, this paper is
selfcontained. In the first part of the next section we briefly review
the quantization of the abelian gauge field theories in the Feynman
gauge discussed in refs. \ferqed\ and \ferabe.
We consider small quantum perturbations $A^{\rm qu}$ around a
topologically nontrivial solution of the classical equations of motion
$A^{\rm I}$.
In order to evaluate the generating functional of the amplitudes between
the fermionic currents of the Schwinger model, we integrate the path
integral over the gauge degrees of freedom.
In this way we obtain the nonlocal GTM mentioned above as an effective
field for the electrons on a Riemann surface.
It is important to stress at this point that the inverse procedure, in
which one integrates first over the fermionic fields, seems to be not
very convenient on curved space-times as it is in the flat case. First
of all, in fact, there are complications with the zero modes and the
topologically nontrivial solutions of the equations of motion $A^{\rm I}$.
Secondly, also if we succeed in the integration, as a result we should
obtain, in
analogy with the flat case, a massive theory of vector fields
\ref\bnd{J. Barcelos-Neto, A. Das, {\it Phys. Rev.} {\bf
D33} (1986), 2262; {\it Zeit. Phys.} {\bf C32} (1986), 527.},
\ref\leutwyler{ H. Leutwyler, {\it Phys. Lett} {\bf 153B}
(1985), 65.}.
Massive field theories are
not easy to handle on a Riemann surface because, for example,
the propagators of massive field theories are still unknown.\smallskip
In the second part of section 2, we perform the remaining integration
over the fermionic degrees of freedom. The advantage of our method is
that now we can use previous experience on the Thirring model \fp,
\ref\sommerfield{C. M. Sommerfield, {\it Ann. Phys.} {\bf 26} (1963),
1.}, \ref\dm{A. Das, V. S. Mathur, {\it Phys. Rev.} {\bf D33} (1986),
489.} and apply it to the GTM. Nevertheless, a crucial point in order to
solve the GTM is the
calculation of the chiral determinant of free fermions in the presence
of an external gauge fields $B$. Unfortunately,
this calculation becomes involved when $B$ is topologically nontrivial,
so we assume here that $B$ belongs to a trivial line bundle on the
Riemann surface. Only in this case, in fact, we can compute the
determinant using the explicit formula of ref. \fp.
For this reason, we have to set $A^{\rm I}=0$ in the
generating functional of the Schwinger model.
The final computation of the generating functional of the correlation
functions between the fermionic current is achieved using slight
generalizations of the formulas obtained in ref. \sommerfield.
The details of the calculation are given in appendix A (generating
functional) and appendix B (partition function).
\vfill\eject
\newsec{INTEGRABILITY OF THE SCHWINGER MODEL ON A RIEMANN SURFACE}
\vskip 1cm
We consider here the Schwinger model \sch\
defined by the following action:
\eqn\action{S=\int_Md^2x\sqrt{g}\left[-{1\over
4}F^{\mu\nu}F_{\mu\nu}+i\bar\psi\gamma^\mu(x)
(D_\mu-ieA_\mu)\psi+eB_\mu\bar\psi\gamma^\mu(x)\psi\right]}
In eq. \action\ $e$ is the coupling constant and $M$ is a closed and
orientable Riemann surface of genus $g$ provided with an Euclidean
conformally flat metric $g_{\mu\nu}$, $\mu,\nu=0,1$. $g_{\mu\nu}$ is induced
by the distance:
\eqn\realmetr{ds^2=e^{-h(x_0,x_1)}(dx_0^2+dx_1^2)}
Together with the metric we need also the zweibein $e^\alpha_\mu$,
$\alpha,\beta=0,1$ being the frame indices. The zweibein is defined so that:
\eqn\zweibein{g^{\mu\nu}(x)=e_\alpha^\mu e_\beta^\nu\eta^{\alpha\beta}}
where $\eta^{\alpha\beta}={\rm diag}(1,1)$ is the flat metric in the
local frame. From this definition it descends that $g\equiv{\rm
det}g_{\mu\nu}=({\rm det} e_\mu^\alpha)^2$.
Moreover, $\gamma^\mu(x)\equiv e^\mu_\alpha(x)\gamma^\alpha$, the
$\gamma^\alpha$ denoting the usual $\gamma-$matrices expressed in terms of
the Pauli matrices $\sigma_1$ and $\sigma_2$ as follows:
$$\gamma^0=-\sigma_2\qquad\qquad\qquad \gamma^1=\sigma_1$$
Finally, we have defined
$F_{\mu\nu}=\partial_\mu A_\nu-\partial_\nu A_\mu$
and $\bar\psi=\psi^\dagger\gamma^0$.\smallskip
Due to the fact that on a Riemann surface the fermionic fields admit
spin structures, we suppose that $\bar \psi$ and $\psi$ transform
according to the following rules when transported along the nontrivial
homology cycles $A_i$, $i=1,\ldots,g$, of the Riemann surface $M$:
$$\bar\psi\rightarrow e^{-2\pi i u_i}\bar \psi\qquad\qquad\qquad
\psi\rightarrow e^{2\pi i u_i}\bar \psi$$
Along the homology cycles $B_i$, we have instead:
$$\bar\psi\rightarrow e^{-2\pi i v_i}\bar \psi\qquad\qquad\qquad
\psi\rightarrow e^{2\pi i v_i}\bar \psi$$
$A_i$ and $B_i$ are chosen here in such a way that they for a canonical
basis for the nontrivial homology cycles
\ref\amv{L. Alvarez-Gaum\'e,
G. Moore, C. Vafa, {\it Comm. Math.
Phys.} {\bf 106} (1986), 1.}.
We notice also that the covariant derivative $D_\mu$ for the
fermions appearing in eq. \action\ is given by:
$$D_\mu=\partial_\mu+\omega_\mu$$
where $\omega_\mu={1\over
8}e_{\alpha\nu}\nabla_\mu e^\nu_\beta[\gamma^\alpha,\gamma^\beta]$ and
$\nabla_\mu$ is the covariant derivative acting on the zweibein
$e_\beta^\nu$.
However, in two dimensions the spin field $\omega_\mu$ does not
contribute to the action \action, so that we will set
$D_\mu\equiv\partial_\mu$ in the rest of this work.\smallskip
Here we are interested in computing the partition function of the
Schwinger model and the amplitudes of the currents:
\eqn\currents{j^\mu(x)=[\bar\psi\gamma^\mu\psi](x)}
In eq. \action,
the external source of these currents is $B_\mu$.\smallskip
In order to fix the abelian gauge symmetry of eq. \action\ and quantize
the theory, we impose the covariant gauge:
\eqn\covgauge{\partial_\mu A^\mu=0}
As a consequence, we have to solve the path integral:
\eqn\pimain{Z[B_\mu]=\int DA_\mu D\bar\psi D\psi e^{-(S+S_{gf})}}
where the gauge fixing term $S_{gf}$ is given by:
\eqn\sgf{S_{gf}={1\over 2\lambda}\int_M d^2x\sqrt{g}(\partial_\mu
A^\mu)^2}
We ignore the Faddeev-Popov term, since the ghost are decoupled
from all the other fields in the abelian case we are treating.
Moreover, we will choose the Feynman gauge, putting $\lambda=1$ in eq.
\sgf.\smallskip
We start to evaluate eq. \pimain\ integrating over the gauge fields.
Therefore, we need to compute the following path integral:
\eqn\pigauge{
Z[j^\mu]=\int DA_\mu
e^{
\int_Md^2x
\sqrt{g}
\left[-
{
1\over
4
}
F_{\mu\nu}
F^{\mu\nu}
+ej^\mu A_\mu+
{
1\over
2
}
(\partial_\mu A^\mu)^2\right]
}
}
To this purpose, it is better to use complex coordinates
$z=x+ix_0$ and $\bar z=x_0-ix_1$.
In the new system of coordinates, the metric is given by:
\eqn\cmplmetr{g_{z\bar z}=g_{\bar z z}=e^{-h(z,\bar
z)}\qquad\qquad\qquad g_{zz}=g_{\bar z\bar z}=0\qquad\qquad\qquad
g^{z\bar z}g_{z\bar z}=1}
It is also convenient to decompose the gauge fields in the following way:
\eqn\decomp{A_z=A^{\rm I}_z+A_z^{\rm qu}+A_z^{\rm har}\qquad\qquad\qquad
A_{\bar z}=A^{\rm I}_{\bar z}+A_{\bar z}^{\rm qu}+A_{\bar z}^{\rm har}}
where $A^{\rm I}_z$ and $A^{\rm I}_{\bar z}$ are instantonic solutions
of the equations of motion in the Feynman gauge corresponding to
$c_1=k\in${\bf Z}. Here $c_1$ denotes the first Chern class:
$$c_1={1\over 2\pi}\int_Md^2zF_{z\bar z}(A^{\rm I})$$
The explicit form of the fields $A^{\rm I}$ has been given in ref. \ferabe\
and we do not report it.\smallskip
Let us now return to the equations \decomp.
The fields $A_z^{\rm qu}$ and $A_{\bar z}^{\rm qu}$ denote a small
quantum perturbation around the instantonic solutions discussed above.
They contain the transverse and longitudinal gauge degrees of freedom,
which can be represented as a coexact and exact form respectively using
the Hodge decomposition
\ferqed, \fp.
Finally, $A^{\rm har}_z$ and $A^{\rm har}_{\bar z}$ describe the
harmonic components of the gauge fields.
They can be written in the following way:
\eqn\harmpart{A_{\bar z}^{\rm har}d\bar z=
2\pi i({\bf \phi}+\bar\Omega{\bf \theta})\cdot(\Omega-\bar
\Omega)^{-1} \cdot \bar \omega(\bar z)d\bar z}
$A^{\rm har}_zdz$ can be computed from eq. \harmpart\ by complex
conjugation.
$\phi_i$ and $\theta_i$ are real numbers and $\Omega_{ij}$,
$i,j=1,\ldots,g$, is the period matrix which can be computed in terms of
the harmonic differentials $\omega_i(z)dz$:
$$\Omega_{ij}=\oint_{B_i}\omega_j(z)dz$$
As a consequence, it is easy to see that the path integral \pigauge\
becomes in complex coordinates:
\eqn\cmplpi{Z[j^\mu]=\int DA_z^{\rm qu}DA_{\bar z}^{\rm qu}\prod_{i=1}^g
d\theta_id\phi_ie^{-(S^{\rm qu}+S^{\rm I})}}
with
\eqn\squ{S^{\rm qu}=\int_Md^2z\left[g^{z\bar z}\left(\partial_zA^{\rm
qu}_{\bar z}\partial_zA^{\rm qu}_{\bar z}+\partial_{\bar z}A^{\rm
qu}_z\partial_{\bar z}A^{\rm qu}_z
\right)+ej_zA^{\rm qu}_{\bar z}+ej_{\bar z}A^{\rm
qu}_z\right]}
\eqn\sist{S^{\rm I}=e\int_Md^2z\left[j_z(A^{\rm I}_{\bar z}+A^{\rm
har}_{\bar z})+j_{\bar z}(A^{\rm I}_z+A^{\rm har}_z)\right]}
Here we have used the following notation: $d^2z\equiv{dz\wedge d\bar
z\over 2i}$.
Again, since at the end we will put $A^{\rm I}=0$, we have not shown
that the instantonic gauge fields decouple from the action of the
quantum fields $S^{\rm qu}$. A proof is given in ref. \ferabe.
Finally, in complex coordinates the current $j_\mu$ assumes the form:
$$j_z=\bar\psi_\theta\psi_\theta\qquad\qquad\qquad j_{\bar
z}=\bar\psi_{\bar \theta}\psi_{\bar \theta}$$
where $\theta$ and $\bar\theta$ are spinor indices.
In order to compute the path integral \cmplpi, we still need the
propagator $G_{\mu\nu}(z,w)=\langle A_\mu^{\rm qu}(z,\bar z)A_\nu^{\rm
qu}(w,\bar w)\rangle$ of the quantum fields.
The components of the propagator were already derived in refs. \ferqed\
and \ferabe, so that we will report here only the result:
\eqn\gzw{G_{zw}(z,w)=-\int_Md^2tg_{t\bar t}\partial_zK(z,t)\partial_wK(w,t)}
\eqn\gzbwb{G_{\bar z\bar w}(z,w)=-\int_Md^2tg_{t\bar t}
\partial_{\bar z}K(z,t)\partial_{\bar w}K(w,t)}
where $K(z,w)$ is the usual scalar Green function defined by the relations:
\eqn\scalone{
\partial_z\partial_{\bar z}K(z,t)=\delta_{z\bar z}^{(2)}(z,t)-
{g_{z\bar z}\over A}\qquad\qquad\qquad A=\int_Md^2zg_{z\bar z}}
\eqn\scaltwo{
\partial_{\bar
z}\partial_tK(z,t)=-\delta^{(2)}_{\bar zt}(z,t)+\sum\limits_{i,j=1}^g
\bar\omega_i(\bar z)\left[{\rm Im}\enskip \Omega\right]^{-1}_{ij}\omega_j(t)}
\eqn\scalthree{\int_Md^2t g_{t\bar t}K(z,t)=0}
One of the characteristics of the propagator given in eqs.
\gzw\ and \gzbwb\ that will be important in the following is that it is
orthogonal with respect to the zero modes.
This can be seen for example from the fact that inserting $G_{zw}(z,w)$
in the equations of motion coming from the action \squ, we get:
\eqn\eqmot{\partial_{\bar z}g^{z\bar
z}\partial_{\bar z}G_{zw}(z,w)=\delta^{(2)}_{z \bar z}(z,w)-
\sum\limits_{i,j=1}^g\bar\omega_i(\bar
z)\left[{\rm Im}\enskip \Omega\right]^{-1}_{ij}\omega_j(w)}
An analogous equation can be derived for the component $G_{\bar z\bar
w}(z,w)$ of the propagator.
It is important to notice that in the right hand side of eq. \eqmot\
the second term is the projector
on the space of the zero modes.
Moreover, if we consider an external current $J_{\bar t}(t,\bar t)$ and
decompose it as follows:
\eqn\currdec{J_{\bar t}(t,\bar t)=\partial_{\bar t}\chi(t,\bar t)+i\bar a_i
[Im\enskip
\Omega]^{-1}_{ij}\bar\omega_j(\bar t)}
with
$$\bar a_i=i\int d^2t\omega_i(t)J_{\bar t}(t,\bar t)$$
we have:
\eqn\orthog{\partial_{\bar z} g^{z\bar z}\partial_{\bar z}\int
d^2t G_{zt}(z,t) J_{\bar t}(t,\bar t)=\partial_{\bar z} \bar\chi(z,\bar z)}
A similar formula is valid for the component $G_{\bar z\bar w}(z,w)$:
\eqn\ccorthog{\partial_z g^{z\bar z}\partial_z\int d^2t G_{\bar z \bar
t}(z,t) J_t(t,\bar t)=\partial_z \chi(z,\bar z)}
showing the orthogonality of the propagator with respect to the harmonic
components of the gauge fields.\smallskip
Now we notice that in eq. \cmplpi\ the integral over the fields $A^{\rm
qu}$ is gaussian.
Therefore, it is possible to eliminate the terms containing the interaction
between the current $j^\mu$ and the fields $A^{\rm qu}$ performing a
shift of the fields. To this purpose, we define new fields:
$$A_z^{\prime {\rm qu}}=A_z^{\rm qu}+
{e\over 2}\int d^2w
G_{zw}(z,w)J_{\bar w}(w,\bar w)$$
$$A_{\bar z}^{\prime {\rm qu}}=A_{\bar z}^{\rm qu}+
{e\over 2}\int d^2w
G_{\bar z\bar w}(z,w)J_w(w,\bar w)$$
Inserting the new fields $A^{\prime{\rm qu}}$ in the action $S^{\rm
qu}$ given in eq. \squ\ we yield:
$$S^{\rm qu}=
\int_Md^2z
\left[
g^{z\bar z}
\partial_{\bar z}
A^{\prime{\rm qu}}_z
\partial_{\bar z}
A^{\prime{\rm qu}}_z
-
{e^2\over 2}
j_{\bar z}
(z,\bar z)\int_Md^2w
G_{zw}(z,w)
j_{\bar w}
(w,\bar w)+
\right.$$
\eqn\squprime
{
\left.
\int_Md^2zg^{z\bar z}
\partial_z
A^{\prime{\rm qu}}_{\bar z}
\partial_z
A^{\prime{\rm qu}}_{\bar z}
-
{e^2\over 2}
j_z(z,\bar z)\int_Md^2w
G_{\bar z\bar w}(z,w)
j_w(w,\bar w)+\right]}
We notice that we have obtained eq. \squprime\ integrating by part in eq.
\squ. This is allowed because the fields $A^{\rm qu}$ and $A^{\prime{\rm
qu}}$ are orthogonal with
respect to the harmonic part. The $A^{\rm qu}$ are orthogonal by
definition, while the fields $A^{\prime {\rm
qu}}_z$ and $A^{\prime {\rm qu}}_{\bar z}$ do not contain any harmonic
component due to the properties of orthogonality \orthog\ and \ccorthog\
of the propagator.
As a consequence, the operators $\partial_z$ and $\partial_{\bar z}$
in \squ\ act on a space of $(0,1)$ and $(1,0)$ forms respectively which
is orthogonal to the harmonic sector and we are free to integrate by
parts in eq. \squ.
Using the action \squprime\ in eq. \cmplpi\ and integrating over the
fields $A^{\prime{\rm qu}}$ we get, apart from a constant factor:
\eqn\newacgf{Z[j^\mu]=\prod\limits_{i=1}^gd\theta_id\phi_ie^{-S_{\rm
I}}e^{W[j]}}
where
\eqn\potgf{W[j]={e^2\over 4}\int_Md^2zd^2w\left[j_{\bar z}(z,\bar
z)G_{zw}(z,w)j_{\bar w}(w,\bar w)+j_z(z,\bar z)G_{\bar z\bar
w}(z,w)j_w(w,\bar w)\right]}
Now we substitute the above result in the path integral of the Schwinger
model \pimain\ after performing the necessary change of coordinates
$(z,\bar z)\rightarrow(x_0,x_1)$.
The upshot is:
\eqn\pieff{Z[B_\mu]=\int D\bar\psi
D\psi\prod\limits_{i=1}^gd\theta_id\phi_ie^{S_{eff}}}
where $S_{eff}$ is the action of the GTM:
$$S_{eff}=
\int_Md^2x\rdc{x}\left[i\bar\psi
\gamma^\mu(x)\left(\partial_\mu-ie(B_\mu+A_\mu^{\rm har}+A_\mu^{\rm
I})\right)\psi+\right.$$
\eqn\seff{\left.{e^2\over 4}\int_Md^2y\rdc{y}j^\mu(x)G_{\mu\nu}(x,y)
j^\nu(y)\right] }
$G_{\mu\nu}(x,y)$ is the propagator of the gauge fields written in real
coordinates.
In terms of the complex components \gzw\ and \gzbwb\ we have:
$$G_{11}(x_0,x_1;y_0,y_1)=-G_{22}(x_0,x_1;y_0,y_1)={\rm
Re}[G_{zw}(z,w)]$$
$$G_{12}(x_0,x_1;y_0,y_1)=G_{21}(x_0,x_1;y_0,y_1)=-{\rm
Im}[G_{zw}(z,w)]$$
In the action \seff, $G_{\mu\nu}(x,y)$ plays the role of a relativistic
potential through which the fermionic currents interact. Thus
$S_{eff}$ describes the effective theory of fermions after removing the
unphysical (in two dimensions) gauge degrees of freedom. The model
associated to the action \seff\ has been already discussed in ref.
\ferabe\ and we only
notice that this effective theory of the electrons can be
viewed as a nonlocal generalization \foot{Local generalizations of
this kind have been studied for example in ref.
as a way for
regularizing the singularities of the gauge field theories. Here the
regularization comes from the fact that the space time is nonflat.} of
the massless Thirring model \thirring.
As we will show, also this generalization is an integrable model.
To verify this, it is convenient to exploit the following formula, which
is an extension to our nonlocal case of the analogous formulas given in
refs. \sommerfield, \dm:
$${\rm exp} \left[-{1\over
4}\int_Md^2xd^2y\rdc{x}\rdc{y}G_{\mu\nu}(x,y){\delta^2\over
\delta B_\mu(x)\delta B_\nu(y)}\right]{\rm
exp}\left[-e\int_Md^2x\rdc{x} j^\mu B_\mu\right]=$$
\eqn\formone{{\rm exp}\int d^2x\rdc{x}\left[-ej^\mu B_\mu-{e^2\over
4}\int_Md^2y\rdc{y}G_{\mu\nu}(x,y) j^\nu(y)j^\mu(x)\right]}
The proof of this equation is straightforward (see appendix A).
As a consequence of \formone, the path integral of the Schwinger model
can be rewritten in this way:
$$Z[B_\mu]={\rm exp} \left[-{1\over
4}\int_Md^2xd^2y\rdc{x}\rdc{y}G_{\mu\nu}(x,y){\delta^2\over
\delta B_\mu(x)\delta B_\nu(y)}\right]$$
\eqn\pia{\int D\bar\psi D\psi \prod\limits_{i=1}^gd\theta_i\phi_i{\rm
exp}\left[
-\int_Md^2x\rdc{x}\left(i\bar\psi\gamma^\mu(x)\partial_\mu\psi+
ej^\mu(B_\mu+A^{\rm har}_\mu+A^{\rm I}_\mu)\right)\right]}
The advantage of having used the formula \formone\ is that now the
action in the fermionic path integral becomes gaussian and it amounts to
a chiral determinant:
\eqn\piferm{{\rm det}\ng=\int D\bar\psi D\psi {\rm exp}\left[-\int_Md^2x
\rdc{x} \left(i\bar\psi\gamma^\mu(x)\partial_\mu\psi+ej^\mu(B_\mu+
A_\mu^{\rm har}+A_\mu^{\rm I})\right)\right]}
where
$\ng=\gamma^\mu(x)\left(\partial_\mu-ie(B_\mu+A_\mu^{\rm har}+A^{\rm
I}_\mu)\right)$.
Unfortunately, at least to our knowledge, the above determinant has not
yet been computed in the presence of
gauge fields $A_\mu^{\rm I}$ belonging to a nontrivial topological
sector.
For this reason, we will set in the following $A^{\rm I}_\mu=0$ in eq.
\piferm.
Moreover, we need also some more informations about the
external gauge field
$B_\mu$.
Here it is convenient to choose $B_\mu$ purely transverse:
$$B_\mu(x)=\epsilon_{\mu\nu}\partial^\nu\varphi$$
where $\varphi$ is a real scalar field.
The motivation of this choice is that in any case the longitudinal
components of $B_\mu$ are physically irrelevant in the path integral
\piferm\ while the harmonic part is already present through the fields
$A^{\rm har}_\mu$.
Looking at eq. \piferm, we can also conclude that the piece containing the
harmonic components $A^{\rm har}_\mu$ factorizes from the rest of the
determinant. As a matter of fact, due to the orthogonality properties of
the Hodge decomposition \fp, we have:
$$\int_Md^2x\rdc{x}j^\mu A_\mu^{\rm har}=\int_Md^2x\rdc{x}j^\mu_{\rm
har}A^{\rm har}_\mu$$
where $j_{\rm har}^\mu$ is the harmonic component of the current
$j^\mu$ of eq. \currdec.
This factorization is evident also from the following formula, derived
in \fp, which gives the explicit form of the determinant \piferm:
\eqn\detdb{{\rm det}\ng={\rm exp}\left[-{e^2\over
2\pi}\int_Md^2x\rdc{x}B_\mu B^\mu\right]\left({{\rm det}({\rm
Im}\Omega)A\over {\rm det}'\triangle_0}\right)^{1\over
2}\left|\theta{\scriptstyle
\left[\matrix{{\bf u}+{\bf \theta}\cr{\bf v}+{\bf
\phi}\cr}\right]}(0,\Omega)\right|^2}
where $A=\int_Md^2x\rdc{x}$ represents the area of $M$ and ${\rm
det}'\triangle_0$ is the regularized determinant (without the constant
zero mode) of the laplacian $\triangle_0$.
Moreover, $\theta{\scriptstyle
\left[\matrix{{\bf u}+{\bf \theta}\cr{\bf v}+{\bf
\phi}\cr}\right]}(0,\Omega)$ is a theta function
\ref\fay{J. D. Fay, {\it Lect. Notes in Math.} {\bf 352}, Springer
Verlag, 1973.} of periods $u_i+\theta_i$ along the cycles $A_i$ and
periods $v_i+\phi_i$ along the cycles $B_i$.
Inserting the determinant \detdb\ in eq. \pia\ we get:
$$Z[B_\mu]=C{\rm exp}\left[-{1\over 4}\int_Md^2xd^2y
\rdc{x}\rdc{y}G_{\mu\nu}(x,y){\delta^2\over\delta B_\mu(x)\delta
B_\nu(y)}\right]\times$$
\eqn\pib{{\rm exp}\left[-{e^2\over
2\pi}\int_Md^2x\rdc{x}B_\mu(x)B^\mu(x)\right]}
with
$$C=
\left({{\rm det}({\rm
Im}\Omega)A\over {\rm det}'\triangle_0}\right)^{1\over
2}\left|\theta{\scriptstyle
\left[\matrix{{\bf u}+{\bf \theta}\cr{\bf v}+{\bf
\phi}\cr}\right]}(0,\Omega)\right|^2$$
In order to compute $Z[B_\mu]$, we use a slight generalization of the
formulas contained in \sommerfield\ showing, see appendix A, that the
generating functional satisfies the following equation\foot{ The symbol
$z$ used in the formula below and in the following is not a complex
variable, but represents the two real variables $z_0,z_1\in M$.}:
\eqn\functeq{\int_Md^2x\rdc{x}\left(\delta^k_{\phantom{k}\mu}
\delta^{(2)}(z,x)-{e^2\over 2\pi}G^k_{\phantom{k}\mu}(z,x)\right)
{\delta\over\delta B_\mu(x)}Z[B_\mu]=-{e^2\over \pi}A^k(z)Z[B_\mu]}
To extract the explicit form of the
generating functional $Z[B_\mu]$ from the above equation,
we need a kernel $M^\rho_{\phantom{\rho}k}(y,z)$ with the property:
\eqn\mrkdef{\int_Md^2z\rdc{z}M^\rho_{\phantom{\rho}k}(y,z)
\left(\delta^k_{\phantom{k}\mu}
\delta^{(2)}(z,x)-{e^2\over
2\pi}G^k_{\phantom{k}\mu}(z,x)\right)=\delta^\rho_{\phantom{\rho}\mu}
\delta^{(2)}(y,x)}
Eq. \mrkdef\ can be solved by iteration supposing that $\left|{e^2\over
2\pi}\right |<<1$:
$$M^\rho_{\phantom{\rho}k}(y,z)=\delta^\rho_{\phantom{\rho}k}(y,z)
\delta^{(2)}(y,z)+{e^2\over 2\pi}G^\rho_{\phantom{\rho}k}(y,z)+$$
\eqn\mrok{\sum\limits_{n=2}^\infty\left({e^2\over
2\pi}\right)^n\int_Md^2x_1\rdc{x_1}\ldots d^2x_{n-1}\rdc{x_{n-1}}
G^\rho_{\phantom{\rho}\mu_1}(y,x_1)
G^{\mu_1}_{\phantom{\rho}\mu_2}(x_1,x_2)\ldots
G^{\mu_{n-1}}_{\phantom{\rho}k}(x_{n-1},z)}
Moreover, exploiting the fact that $G_{\mu\nu}(x,y)=G_{\nu\mu}(y,x)$, we
can easily see that the following equation holds (see appendix A):
\eqn\propmrk{M^{\rho k}(y,z)=M^{k\rho}(z,y)}
Applying to both sides of eq. \functeq\ the operator $\rdc{z}
M^\rho_{\phantom{\rho}k}(y,z)$ and integrating over $z$, we arrive at
the following expression for $Z[B_\mu]$:
\eqn\derivazb{{\delta\over\delta B_\rho(y)}Z[B_\mu]=-{e^2\over
\pi}\int_Md^2z\rdc{z}
M^\rho_{\phantom{\rho}k}(y,z)B^{k}(z)Z[B_\mu]}
It is now easy to verify using the property \propmrk\ of the kernel
$M^\rho_{\phantom{\rho}k}(y,z)$ that the solution of the above equation
is given by:
\eqn\genfunct{Z[B_\mu]=Z_0{\rm exp}\left[-{e^2\over
2\pi}\int_Md^2zd^2y\rdc{z}\rdc{y} B_\rho(y)M^{\rho k}(y,z)B_k(z)\right]}
This formula is the explicit expression of
the generating functional $Z[B_\mu]$. Deriving eq. \genfunct\ in $B_\mu$,
we are able to compute
the amplitudes between the
currents \currents. Still we have to derive the partition function
$Z_0$.
The explicit calculation is performed in appendix B and here we just state
the result:
\eqn\partfunct{Z_0={\rm exp}\left[-{e^2\over\pi}{\rm Tr}\enskip{\rm log}
\left(M^\rho_{\phantom{\rho}k}(y,z)\right)\right]}
where
$$-{\rm Tr}\enskip{\rm log}
\left(M^\rho_{\phantom{\rho}k}(y,z)\right)={1\over
4}\int_Md^2x\rdc{x} G^k_{\phantom{k}k}(x,x)+$$
$${1\over \pi}\left({e\over
4}\right)^2 \int_Md^2xd^2z\rdc{x}\rdc{z}G^k_{\phantom{k}\rho}(z,x)
G^\rho_{\phantom{\rho}k}(x,z)+
{1\over 4}\sum\limits_{n=2}^\infty{1\over 4(n+1)}
\left({e^2\over
2\pi}\right)^n\times$$
\eqn\trlogmrk{\int_Md^2x\rdc{x}
d^2z\rdc{z}
d^2x_1\rdc{x_1}\ldots d^2x_{n-1}\rdc{x_{n-1}}
G^\rho_{\phantom{\rho}\mu_1}(x,x_1)\ldots
G^{\mu_{n-1}}_{\phantom{\mu_n}k}(x_{n-1},z)
G^k_{\phantom{k}\rho}(z,x)}
\vskip 1cm
\newsec{CONCLUSIONS}
\vskip 1cm
In this paper we have shown that the Schwinger model on a Riemann
surface is integrable in absence of topologically nontrivial gauge
fields.
The generating functional and the partition function are given in eqs.
\genfunct\ and \partfunct\ respectively.
Many questions and possible developments remain open. For example, it
remains the problem of integrability in the case of nontrivial line
bundles.
In fact, due to the presence of zero modes, it is difficult to compute
the chiral determinant \piferm\ explicitly when $A^{\rm I}_\mu\ne 0$.
In this respect, one of the advantages of our procedure is that,
integrating over the gauge fields in the path integral of the Schwinger
model, we obtain the GTM, from which one can read many of the
properties of QED$_2$, like the high energy behavior discussed above and
investigated in ref. \ferabe, which exist also in the presence of
topologically nontrivial gauge fields. It is possible to extract these
properties
from the relativistic potential $G_{\mu\nu}(x,y)$ governing the
electromagnetic interactions and whose components are explicitly given
in eqs. \gzw, \gzbwb.\smallskip
It would also be interesting to see if the GMT defined in eqs. \pieff\
and \seff\ can be solved using the inverse scattering method \ref\ism{L.
D. Faddeev, Integrable models in $1+1$ dimensional quantum field theory.
In: Les Houches, Session XXXIX, 1982, Recent Advances in Field Theory
and Statistical Mechanics, J.-B. Zuber, R. Stora (editors), 553-608.
Elsevier Science Publishers 1984.}.
The nonlocality of the model should have nontrivial consequences on the
form of the classical $r-$matrix.\smallskip
Finally, we stress the fact that two dimensional models quantized on
Riemann surfaces represent an important tool in order to study the effects of
an external gravitational background on the dynamics.
For example, a Riemann surface with punctures which is imbedded in three
dimensions has a topology which is similar to that of a wormhole
\ref\fercmp{F. Ferrari, {\it Comm. Math. Phys.} {\bf 156} (1993), 179.}.
Moreover, the Riemann surfaces, having a very rich structure,
are also very suitable to investigate the possibility of
creating quantum mechanical states in curved space-times.
As it is well known, in fact, in the canonical quantization of field theories
new quantum mechanical states can be generated
due to the presence of diffeomorphisms which are not deformable to
the identity. In four dimensions these states have been
called geons \ref\geons{J.
L. Friedman, R. D. Sorkin, {\it Phys. Rev. Lett.} {\bf 44} (1980), 1100;
C. J. Isham, {\it Phys. Lett.} {\bf 196B} (1981), 188;
J. L. Friedman, D. M. Witt, {\it Phys. Lett.} {\bf 120B} (1983), 324.}.
In two dimensions, the Riemann surfaces with crystallographic group of
symmetry provide an excellent way of reproducing this effect.
Moreover, the advantage of working in two dimensions is that one can also
explicitly construct the quantum mechanical states induced by the topology.
This has been done in refs. \fercmp\ and \ref\ferijmpa{F. Ferrari, {\it
Multivalued Fields in the Complex Plane and Braid Group Statistics},
Preprint LMU-TPW 92-24, to be published in {\it Int. Jour. Mod. Phys.}
{\bf A}.} for some simple conformal field theories and considering the
crystallographic groups $Z_n$ and $D_n$.
We hope that the explicit form of the partition function and of the
generating functional computed here will allow the extensions of these
results to the more physical Schwinger model.
\vskip 1cm
\centerline{\bf ACKNOWLEDGEMENTS}
\vskip 1cm
The author would like to thank M. B. Green, C. M. Hull, M. Mintchev,
S. Theisen and J. Wess for many interesting discussions and for their
interest in this work.
This work was supported by a grant of Consiglio Nazionale delle
Ricerche, P. le A. Moro 7, Roma.
\vskip 1cm
\appendix{A}{}
\vskip 1cm
First of all, we prove the formula \formone:
$${\rm exp} \left[\alpha
\int_Md^2xd^2y\rdc{x}\rdc{y}G_{\mu\nu}(x,y){\delta^2\over
\delta B_\mu(x)\delta B_\nu(y)}\right]{\rm
exp}\left[\beta\int_Md^2u\rdc{u} j^\sigma(u) B_\sigma(u)\right]=$$
\eqn\gformone{
\sum\limits_{n=0}^\infty{1\over n!}\left[\alpha\int_Md^2xd^2y\rdc{x}\rdc{y}
G_{\mu\nu}(x,y){\delta^2\over\delta B_\mu(x)\delta
B_\nu(y)}\right]^n{\rm exp}\left[\beta\int_M d^2u\rdc{u}j^\sigma
B_\sigma\right]}
Using now the fact that:
$$\left[\int_Md^2xd^2y\rdc{x}\rdc{y}G_{\mu\nu}(x,y){\delta^2\over
\delta B_\mu(x)\delta B_\nu(y)}\right]
{\rm exp}\left[\beta\int_M d^2u\rdc{u}j^\sigma
B_\sigma\right]=$$
$$\beta^2
\int_Md^2xd^2y\rdc{x}\rdc{y}G_{\mu\nu}(x,y)j^\mu(x)j^\nu(y)$$
and applying it to eq. \gformone, we obtain:
$${\rm exp} \left[\alpha
\int_Md^2xd^2y\rdc{x}\rdc{y}G_{\mu\nu}(x,y){\delta^2\over
\delta B_\mu(x)\delta B_\nu(y)}\right]{\rm
exp}\left[\beta\int_Md^2u\rdc{u} j^\sigma(u) B_\sigma(u)\right]=$$
$${\rm exp}\left[\alpha\beta^2
\int_Md^2xd^2y\rdc{x}\rdc{y}G_{\mu\nu}(x,y)j^\mu(x)j^\nu(y)\right]
{\rm exp}\left[\beta\int_M d^2u\rdc{u}j^\sigma
B_\sigma\right]$$
Setting $\alpha=-{1\over 4}$ and $\beta=-e$, one recovers exactly eq.
\formone.
\smallskip
Now, we derive eq. \functeq\ using a slight generalization
of the functional formulas of \sommerfield.
Setting $C=1$ in eq. \pib, we consider the functional ${\delta
Z[B_\mu]\over \delta B_k(z)}$:
$${\delta
Z[B_\mu]\over \delta B_k(z)}=-{e^2\over \pi}{\rm exp}\left[-{1\over
4}\int_Md^2xd^2y\rdc{x}\rdc{y}G_{\mu\nu}(x,y){\delta^2\over\delta
B_\mu(x)\delta B_\nu(y)}\right]\times$$
\eqn\zbmder{B^k(z){\rm exp}\left[-{e^2\over
2\pi}\int_Md^2x\rdc{x}g^{\mu\nu} B_\mu B_\nu\right]}
To simplify this equation, we have first to introduce the operators
$$S(\beta)={\rm exp}\left[-\beta\int_Md^2xd^2y\rdc{x}\rdc{y}
G_{\mu\nu}(x,y){\delta^2\over \delta B_\mu(x)\delta B_\nu(y)}\right]$$
and $S^{-1}(\beta)=S(-\beta)$, where $\beta$ is an arbitrary parameter.
Now we compute:
$${d\over d\beta}\left[S(\beta)B^k(z)S^{-1}(\beta)
\right]=$$
\eqn\bcomm{-S(\beta)\left[\int_Md^2xd^2y\rdc{x}\rdc{y}G_{\mu\nu}(x,y)
{\delta^2\over \delta B_\mu(x)\delta
B_\nu(y)},B^k(z)\right]S^{-1}(\beta)}
where $[,]$ denotes the usual commutator.
The derivation of the commutator contained in eq. \bcomm\ is simple and
yields:
$$\left[\int_Md^2xd^2y\rdc{x}\rdc{y}G_{\mu\nu}(x,y)
{\delta^2\over \delta B_\mu(x)\delta
B_\nu(y)},B^k(z)\right]=2\int_Md^2x\rdc{x}G^k_{\phantom{k}\mu}(z,x)
{\delta\over \delta B_\mu(x)}$$
Therefore, using the fact that $S(\beta)S^{-1}(\beta)=1$, eq. \bcomm\
becomes:
$${d\over d\beta}\left[S(\beta)B^k(z)S^{-1}(\beta)
\right]=-2\int_Md^2x\rdc{x}G^k_{\phantom{k}\mu}(z,x)
{\delta\over \delta B_\mu(x)}$$
Integrating this equation in $\beta$ between $0$ and $1\over 4$, we get:
$$[S(
{1\over 4}
),B^k(z)]=-
{1\over 2}
\int_Md^2x\rdc{x}
G^k_{\phantom{k}\mu}
(z,x){\delta\over \delta B_\mu(x)}
S({1\over 4})$$
Applying this equation to eq. \zbmder, we have:
$${\delta
Z[B_\mu]\over \delta B_k(z)}=-{e^2\over \pi}\left[B^k(z)-{1\over
2}\int_Md^2x\rdc{x} G^k_{\phantom{k}\mu}(z,x){\delta\over \delta
B_\mu(x)} \right]Z[B_\mu]$$
which is clearly equivalent to eq. \functeq\ after setting:
$${\delta Z[B_\mu]\over \delta
B_k(z)}=\int_Md^2x\rdc{x}\delta^k_{\phantom{k}\mu}
\delta^{(2)}(z,x){\delta Z[B_\mu]\over \delta B_\mu(x)}$$
Finally, we verify eq. \propmrk.
To this purpose, we exploit the explicit form of
$M^\rho_{\phantom{\rho}k}(y,z)$ given in eq. \mrok.
At the zeroth and first order in $e^2\over 2\pi$, the proof of eq.
\propmrk\ is trivial.
At the $n-$th order, $n\ge 2$, we use the property
$G^\mu_{\phantom{\mu}\nu}(x,y)=
G_\nu^{\phantom{\nu}\mu}(y,x)$ in order to rewrite the $n-$th element of
$M^{\rho k}(y,z)$ as follows:
$$\int_Md^2x_1\ldots d^2x_{n-1}\rdc{x_1}\ldots\rdc{x_{n-1}}
G^\rho_{\phantom{\rho}\mu_1}(y,x_1)G^{\mu_1}_{\phantom{\mu_1}\mu_2}(x_1,x_2)
\ldots G^{\mu_{n-1}k}(x_{n-1},z)=$$
$$\int_Md^2x_1\ldots d^2x_{n-1}\rdc{x_1}\ldots\rdc{x_{n-1}}
G_{\mu_1}^{\phantom{\mu_1}\rho}(x_1,y)
G^{\mu_1}_{\phantom{\mu_1}\mu_2}(x_1,x_2)\ldots$$
\eqn\interm{G^{\mu_{n-2}}_{\phantom{\mu_{n-2}}\mu_{n-1}}(x_{n-2},x_{n-1})
G^{k\mu_{n-1}}(z,x_{n-1})}
Now, we lower and rise the summed indices $\mu_1\ldots\mu_{n-1}$ in eq.
\interm\ as follows:
$$G_{\mu_1}^{\phantom{\mu_1}\rho}(x_1,y)\rightarrow
G^{\mu_1\rho}(x_1,y)\qquad\qquad
G^{\mu_i}_{\phantom{\mu_i}\mu_{i+1}}(x_i,x_{i+1})\rightarrow
G_{\mu_i}^{\phantom{\mu_i}\mu_{i+1}}(x_i,x_{i+1})$$
$$G^{k\mu_{n-1}}(z,x_{n-1})\rightarrow
G^k_{\phantom{k}\mu_{n-1}}(z,x_{n-1})$$
Finally, we change the name of the variables setting $x_{n-i}=x_i$ and
$x_i=x_{n-i}$.
Combining these two actions together, we have:
$$\left. M^{\rho k}(y,z)\right|_{n-{\rm th}\atop {\rm order}}=
\int_Md^2x_1\ldots d^2x_{n-1}\rdc{x_1}\ldots\rdc{x_{n-1}}
G^k_{\phantom{k}\mu_1}(z,x_1)
G^{\mu_1}_{\phantom{\mu_1}\mu_2}(x_1,x_2)\ldots$$
$$G^{\mu_{n-2}}_{\phantom{\mu_{n-2}}\mu_{n-1}}(x_{n-2},x_{n-1})
G^{\mu_{n-1}\rho}(x_{n-1},z)\equiv \left. M^{k\rho}(z,y)
\right|_{n-{\rm th}\atop {\rm order}}$$
This concludes our proof of eq. \propmrk.
\vskip 1cm
\appendix{B}{}
\vskip 1cm
In this appendix we derive the partition fuunction $Z_0$ given in
eq. \partfunct.
To this purpose (see \sommerfield), we consider the following
functional:
$$Z_0(\beta)={\rm exp}\left[-\beta\int_Md^2xd^2y\rdc{x}\rdc{y}
G_{\mu\nu}(x,y){\delta^2\over \delta B_\mu(x)\delta B_\nu(y)}\right]\times$$
$$\left.
{\rm exp}
\left[-{e^2\over 2\pi}\int_Md^2x\rdc{x}B_\mu B^\mu\right]\right|_{B_\mu=0}$$
After a simple calculation we find:
$${d\over d\beta}Z_0(\beta)=-\left[\int_Md^2xd^2y\rdc{x}\rdc{y}G_{\mu\nu}(x,y)
{\delta\over\delta B_\mu(x)}{\delta Z_0(\beta)\over \delta B_\nu(y)}
\right]_{B_\mu=0}$$
Now we exploit eq. \derivazb\ in order to evaluate the functional $\delta
Z_0(\beta)\over \delta B_\nu(y)$. Substituting the result in the above
equation and recalling the
fact that at the end we have to put $B_\mu=0$ everywhere, we obtain:
\eqn\dbzb{{d\over d\beta}{\rm log}Z_0(\beta)=-{e^2\over \pi}\int_M
d^2xd^2y\rdc{x}\rdc{y} G_{\mu\nu}(x,y)M_\beta^{\nu\mu}(y,x)}
In the above equation $M^\nu_{\phantom{\nu}\mu}(y,x)$ is defined as in
eq. \mrok\ after the substitution ${e^2\over
2\pi}\rightarrow{e^2\beta\over 2\pi}$.
Integrating both sides of eq. \dbzb\ in $\beta$ between $0$ and $1$ we
have:
$$Z_0(\beta=1)=Z_0(\beta=0){\rm exp}\left[-{e^2\over
\pi}\int_Md^2xd^2y\rdc{x}\rdc{y}G_{\mu\nu}(x,y) \int_0^1d\beta
M_\beta^{\nu\mu}(y,x)\right]$$
Since the primitive in $\beta$ of $M_\beta^{\nu\mu}(x,y)$ is zero when
$\beta=0$, it is clear that $Z_0(\beta=0)=1$.
Remembering that $Z_0(\beta=1)\equiv Z_0$ we obtain eq. \trlogmrk.

\listrefs
\bye